\documentclass[12pt,preprint]{aastex}
\slugcomment{Accepted to Astrophys. J.}

\shorttitle{Lightcurve characteristics of GRBs 980425 and 060218}
\shortauthors{Zhang }

\begin{document}


%\title{BROADBAND PROPERTIES OF GRBS 980425 AND 060218 AND COMPARISON
%WITH LONG-LAG, WIDE-PULSE GAMMA-RAY BURSTS}

\title{Broadband lightcurve characteristics of GRBs 980425 and 060218 and comparison
 with long-lag, wide-pulse GRBs}

\author{Fu-Wen Zhang\altaffilmark{1,2,3}}

\altaffiltext{1}{National Astronomical Observatories/ Yunnan
Observatory,
 Chinese Academy of Sciences, P.O. Box 110, Kunming, Yunnan 650011, China;
 fwzhang@ynao.ac.cn}

\altaffiltext{2}{Department of Mathematics and Physics, Guilin
University of Technology, Guilin, Guangxi 541004, China}

\altaffiltext{3}{The Graduate School of the Chinese Academy of
Sciences, P.O. Box 3908, Beijing 100039, China}

\begin{abstract}
It has been recently argued that low-luminosity gamma-ray bursts
(LL-GRBs) are likely a unique GRB population. Here, we present
systematic analysis of the lightcurve characteristics from X-ray to
gamma-ray energy bands for the two prototypical LL-GRBs 980425 and
060218. It is found that both the pulse width ($w$) and the ratio of
the rising width to the decaying width ($r/d$) of theses two bursts
are energy-dependent over a broad energy band. There exists a
significant trend that the pulses tend to be narrower and more
symmetry with respect to the higher energy bands for the two events.
Both the X-rays and the gamma-rays follow the same $w - E$ and $r/d
- E$ relations. These facts may indicate that the X-ray emission
tracks the gamma-ray emission and both are likely to be originated
from the same physical mechanism. Their light curves show
significant spectral lags. We calculate the three types of lags with
the pulse peaking time ($t_{peak}$), the pulse centroid time
($t_{cen}$), and the cross-correlation function (CCF). The derived
$t_{peak}$ and $t_{cen}$ are a power-law function of energy. The lag
calculated by CCF is strongly correlated with that derived from
$t_{peak}$. But the lag derived from $t_{cen}$ is less correlated
with that derived from $t_{peak}$ and CCF. The energy dependence of
the lags is shallower at higher energy bands. These characteristics
are well consistent with that observed in typical long-lag,
wide-pulse GRBs, suggesting that GRBs 980425 and 060218 may share
the similar radiation physics with them.

\end{abstract}


\maketitle
% main text
\section{Introduction}

Two nearby gamma-ray bursts (GRBs) 980425 and 060218 are detected
respectively, at the redshifts 0.0085 (Tinney et al. 1998) and
0.0331 (Masetti et al. 2006; Mirabal \& Halpern 2006). The isotropic
luminosities (L$_{iso}$) of GRBs 980425 and 060218 are
$1.21\times10^{47}$ erg s$^{-1}$ (Hakkila et al. 2008) and
$1.2\times10^{47}$ erg s$^{-1}$ (Liang et al. 2006, hereafter L06),
respectively, marking them prominent low-luminosity GRBs (LL-GRBs)
with respect to typical GRBs ($L_{iso}\sim10^{50}-10^{52}$ erg
s$^{-1}$). Both of them are associated with observed supernova of
Type Ic, i.e. GRB 980425/SN 1998bw (Galama et al. 1998) and GRB
060218/SN 2006aj (Masetti et al. 2006, Campana et al. 2006; Pian et
al. 2006).

The nature of these two bursts are highly uncertain. Based on the
high detection rate inferred from these two nearby events, Liang et
al. (2007) proposed that these LL-GRBs might form a unique GRB
population, characterized by high local GRB rate, small beaming
factor, and low luminosity (see also Le \& Dermer 2007; Guetta \&
Della Valle 2007). However, the spectral properties of the prompt
emission for these two events are apparently different. The peak
energy ($E_{p,i}$) of the cosmological rest-frame $\nu f_\nu$
spectrum of GRB 060218 is $4.9\pm0.3$ keV, which is well consistent
with the $E_{p,i}-E_{iso}$ correlation (the so-called
Amati-relation) (Amati et al. 2007). This is reasonable if the
Amati-relation is possibly due to a radiation effect (Liang \& Dai
2004). Furthermore, GRB 060218 roughly complies with the
luminosity-lag relation ($L-\tau$ relation) (Gehrels et al. 2006;
L06) derived from typical GRBs (Norris et al. 2000). These facts
indicate that GRB 060218 is a typical X-ray flash, a soft version of
GRBs (Lamb et al. 2005). However, GRB 980425 is an apparent outlier
with respect to the Amati-relation and the $L-\tau$ relation
(Sazonov et al. 2004; Amati 2006). Ghisellini et al. (2006) argued
that this may be a hard-to-soft spectral evolution effect. These
intriguing observations motivate us to make further analysis on the
emission properties of these two events. We focus on their
lightcurve characteristics in an attempt to determine whether
evidence exists to explain their abnormal luminosities. Their light
curves are composed of a smooth, fast-rise-exponential-decay (FRED)
pulse with significant spectral lag (Sazonov et al. 2004; L06).
Using CGAO/BATSE, BeppoSAX and Swift observations, we obtain their
broad band prompt emissions from X-rays to gamma-rays, which are
presented in \S2. We derive the spectral lag ($\tau$) and the
energy-dependence of pulse-width ($w$) and the ratio of pulse
rise-to-decay ($r/d$) for these two events in \S3. Norris et al.
(2005) (hereafter N05) identified a subgroup of GRBs with long-lag,
wide-pulse in their prompt emission profiles. To further examine
whether they share the same properties with typical long-lag,
wide-pulse GRBs (LLWP-GRBs), we also make a comparison of the
temporal properties of the two bursts with that of the LLWP-GRBs in
\S4. Conclusions and discussions are presented in \S 5.

\section{Data}

GRB 980425 was detected on 1998 April 25.90915 UT with one of the
Wide Field Cameras (WFCs) and the Gamma Ray Burst Monitor (GRBM) on
board BeppoSAX. This burst was also observed with the BATSE
instrument on board the Compton Gamma Ray Observatory (CGRO) at
21:49:08.7 UT (trigger 6707). The X-ray light curves with a temporal
resolution of 1 second in energy bands $2-5$ keV, $5-10$ keV, and
$10-26$ keV observed with WFC are available at ASI Science Data
Center\footnote{http://www.asdc.asi.it or
http://www.asdc.asi.it/grb-wfc} (Vetere et al. 2007). The gamma-ray
light curves observed with CGRO/BATSE are obtained via anonymous ftp
from the CGRO/BATSE
website\footnote{ftp://legacy.gsfc.nasa.gov/compton/data/batse/ascii\_data/64ms/}.
They are available in four energy bands, i.e., $25-50, 50-100,
100-300$, and $>$300 keV, with a temporal resolution of 64 ms. The
backgrounds of these light curves are fitted by a polynomial
expression, and they are obtained from the CGRO Science Support
Center (CGROSSC) at NASA Goddard Space Flight Center through its
public
archives\footnote{http://cossc.gsfc.nasa.gov/docs/cgro/batse/batseburst/sixtyfour
\_ms/bckgnd\_fits.html}. Figure 1 shows the background subtracted
light curves in the X-rays and gamma-rays bands (note that the
signal in the $>$300 keV band is not detected, so, the light curve
in this band is not displayed). All the light curves are shown with
respect to the BATSE trigger time without considering the
propagation delay between the spacecrafts.

GRB 060218 was detected with the Swift Burst Alert Telescope (BAT)
on 2006 February 18.149 UT. Its duration $T_{90}\sim$ 2000 s in the
$15-150$ keV energy band. Swift slewed autonomously to the burst,
and the X-ray telescope (XRT) and the UV/Optical Telescope (UVOT)
started collecting data 159 s after the burst trigger. The early
X-ray emission contains a thermal emission component (Campana et al.
2006). L06 derived the X-ray light curves of the nonthermal emission
in energy bands $0.3-2$ keV, $2-5$ keV and $5-10$ keV by subtracting
the thermal emission component from the XRT data. The BAT trigger of
this event is an image trigger. L06 extracted the gamma-ray emission
light curves in the whole BAT energy band ($15-150$ keV). In our
analysis the lightcurve data are taken from L06, which are shown in
Figure 2.

The light curves of the two events in the gamma-ray energy bands are
a long-lag, long duration single-pulse. N05 made an extensive
analysis on a sample of GRBs with a long-lag, wide-pulse observed by
CGRO/BATSE. In order to make comparison of the two events with these
LLWP-GRBs, we used the data of these bursts from N05.

\section{Energy Dependence of Lightcurve Characteristics}

From Figures 1 and 2, we find that there is an obvious trend, the
pulses become narrower at higher energies and the pulse peaks shift
to later times at lower energies from the X-ray to gamma-ray energy
bands for GRBs 980425 and 060218. We also find that the single-pulse
structure of these two bursts apparent at higher energies becomes
less obvious at lower energies. The loss of pulse structure at lower
energies could be due to, partially by lower signal-to-noise
measurements, and also might be due to, partially by some sort of
faint pulse substructure. While checking the dependence in the
different energy bands are the same or different, here we pay
attention how the pulse width and spectral lag depend on energy over
a broad energy band.

\subsection{Pulse Width and Energy Dependence}

Although the single-pulse structure of the two bursts is less
obvious at lower energies, we still model their light curves in
different energy bands by a single FRED pulse. The pulse profiles of
GRBs are found to be self-similar across energy bands (e.g., Norris
et al. 1996). Kocevski et al. (2003) developed an empirical
expression, which can be used to fit the pulses of GRBs well. This
function can be written as,
\begin{equation}
F(t)=F_{m}(\frac{t+t_{0}}{t_{m}+t_{0}})^{r}[\frac{d}{d+r}+\frac{r}{d+r}
(\frac{t+t_{0}}{t_{m}+t_{0}})^{(r+1)}]^{-\frac{r+d}{r+1}},
\end{equation}
where $t_{m}$ is the time of the maximum flux ($F_{m}$) of the
pulse, $t_{0}$ is the offset time, $r$ and $d$ are the rising and
decaying power-law indices, respectively. We fit all the light
curves of GRB 980425 in the different energy bands with equation (1)
and then measure the values of $w$ and $r/d$. The errors of $w$ and
$r/d$ are derived from the simulations by assuming a normal
distribution of the errors of the fitting parameters. The reported
errors are at $1\sigma$ confidence level. The results are listed in
Table 1.

From Table 1, we find a significant trend that the pulses tend to be
narrower and more close to symmetric at higher energies for GRB
980425. We show $w$ and $r/d$ as functions of $E$ in Figure 3
($left$), where $E$ is the geometric mean of the lower and upper
boundaries of the corresponding energy band, which is adopted
throughout this paper unless otherwise referred to. Apparently both
$w$ and $r/d$ are correlated with $E$. A best fit yields $w\propto
E^{-0.20\pm0.04}$ (Fig. 3 [\emph{left-top}]) and $r/d\propto
E^{0.10\pm0.01}$ (Fig. 3 [\emph{left-bottom}]). The detailed results
of the correlation analysis are listed in Table 2. It is found that
the $r/d - E$ relation of GRB 980425 is well consistent with that
observed in the majority of GRBs (e.g., N05; Peng et al. 2006), but
the power-law index of the $w - E$ relation for this event is
somewhat larger than that ($\sim -0.4$) previously observed in
typical GRBs (e.g., Fenimore et al. 1995; Norris et al. 1996; N05).

Both the relations, $w - E$ ($w\propto E^{-0.31\pm0.03}$) and $r/d -
E$ ($r/d\propto E^{0.10\pm0.03}$) for GRB 060218 are also displayed
in Figure 3 ($right$) and listed in Table 2. We find that this burst
roughly satisfies the same $w - E$ relation observed in typical
GRBs, the index of the $w - E$ relation is also shallower, similar
to that observed in GRB 980425. Note that, the distribution of the
power-law indices for a typical GRB sample has a large dispersion,
the median value is $\sim -0.4$ (see, Jia \& Qin 2005; Peng et al.
2006; Zhang et al. 2007, last one hereafter Z07). Thus, it is
possible that there is no a universal power-law index of the $w - E$
relation. We also find that the energy dependence of $r/d$ for the
burst is consistent with that observed in typical GRBs, but the
value of $r/d$ in the $15-150$ keV band has large error. These
results show that both the X-rays and gamma-rays follow the same $w
- E$ and $r/d - E$ relations for GRBs 980425 and 060218, indicating
that the X-ray emission tracks the gamma-ray emission and thus the
two emission are most likely to originate from the same physical
mechanism. A similar case is also found in GRB 060124 (Romano et al.
2006; Zhang \& Qin 2008).

\subsection{Spectral Lag and Energy Dependence}

The light curves of GRBs 980425 and 060218 shown in Figures 1 and 2
display significant spectral lags, with the pulse peaks shifting to
later time at lower energies, similar to that observed in typical
GRBs by several authors (see e.g., Link et al. 1993; Cheng et al.
1995; Norris et al. 1996, 2000; Band 1997; Wu \& Fenimore 2000;
Hakkila \& Giblin 2004, 2006; Chen et al. 2005; Norris \& Bonnell
2006; Yi et al. 2006; Zhang et al. 2006a, 2006b; Hakkila et al.
2007). By using the fitting pulse data, we can measure the pulse
peak time ($t_{peak}$) of each energy band. The results are also
listed in Table 1. Figure 4 [\emph{left-top}] shows the correlation
between $t_{peak}$ and $E$ for GRB 980425. The best fit to the
correlation yields $t_{peak}\propto E^{-0.35\pm0.04}$. The same
analysis for GRB 060218 ($t_{peak}\propto E^{-0.25\pm0.05}$)
performed by L06 is also displayed in Figure 4 [\emph{right-top}].
The $t_{peak}- E$ relations for these two bursts are listed in Table
2. We find that the indices of the $t_{peak}- E$ relations are
different for the two bursts. The pulse peak lags ($\tau_{peak}$)
are defined as the differences between the pulse peak times in
different energy bands (e.g., N05; Z07; Hakkila et al. 2008). The
values of $\tau_{peak}$ between any pairs of the six light curves of
GRB 980425 can be simply obtained and listed in Table 3.

It is well known that the pulse centroid time ($t_{cen}$) can be
easily measured than the pulse peak time, which is depicted as,
$t_{cen}=\int I(t)tdt/\int I(t)dt$, where $I(t)$ is the pulse
intensity (see Appendix A in N05). In general, $t_{cen}$ can be
directly estimated from the observed lightcurve data (e.g., Norris
et al. 2002; N05). The observed data are discrete, so, we simply
replace the integral equation by a sum one, $t_{cen}=\sum
I(t)t\Delta t/\sum I(t)\Delta t$, where $\Delta t$ is the time bin
of the observed data. Using this equation, we measure
$t_{cen}^{\ast}$\footnote{The symbol $\ast$ represents the value is
estimated directly from the observed data.} in the different energy
bands for GRBs 980425 and 060218. The errors are estimated from
simulations by assuming a normal distribution of the errors of the
observed data. The results are reported in Table 1 as well. From
Table 1, we find that $t_{cen}^{\ast}$ and $E$ are also correlated.
The best fit to the correlation yields $t_{cen}^{\ast}\propto E
^{-0.40\pm0.07}$ for GRB 980425 (Fig. 4 [\emph{left-bottom}]) and
$t_{cen}^{\ast}\propto E ^{-0.15\pm0.03}$ for GRB 060218 (Fig. 4
[\emph{right-bottom}]). Meanwhile, the pulse centroid lags
($\tau_{cen}^{\ast}$) are defined by the differences between the
pulse centroid times in different energy bands, which can be
calculated between any of two energy bands and listed in Table 3 as
well. In addition, for a purpose of comparison, we also calculate
$t_{cen}$ and pulse centroid lags ($\tau_{cen}$) from the fitting
light curves. The results are listed in Tables 1 and 3. We find that
the relation between $t_{cen}$ and $E$ is consistent with that
between $t_{cen}^{\ast}$ and $E$ as reported in Table 2. We also
find that the indices associated with the pulse centroid time and
energy for the two bursts are different.

In addition, the lags calculated with the cross correlation function
(CCF), $\tau_{CCF}$, have been widely adopted by many authors (Band
1997; Norris et al. 2000; Wu \& Fenimore 2000; Hakkila \& Giblin
2004, 2006; Chen et al. 2005; Norris \& Bonnell 2006; Yi et al.
2006, 2008; Zhang et al. 2006a, 2006b; Z07; Peng et al. 2007,
Hakkila et al. 2007). In general, $\tau_{CCF}$ can be calculated
directly from the observed data. However, since the time resolution
of the X-ray light curves of GRB 980425 is very low and different
with that of the gamma-ray light curves, we can not directly use the
observed data to measure all lags between any pairs of the light
curves. Thus, we estimate $\tau_{CCF}$ with the normalized light
curves derived from the pulse fits for this event. To reduce the
uncertainty in the lag measurement, we adopt the same approach as
presented by Hakkila \& Giblin (2006). Thanks to the GRB pulse model
(Norris et al. 1996) which is a time-asymmetric function, and has
the additional degrees of freedom than a quadratic (Wu \& Fenimore
2000) or a cubic (Norris et al. 2000) which can result in a more
accurate CCF fit. This model is used to fit the CCF. The reported
lags are derived by averaging lags obtained from CCF measurements
spanning a range of temporal shifts (typically, 6 trial measurements
are made over a broad range of CCF values in the vicinity of the CCF
peak). The errors of lags are evaluated by the simulations. The
results are also reported in Table 3. The CCF lags of GRB 060218
derived by L06 are available and reported in Table 3 as well.

Then immediately arises a question, whether the values of
$\tau_{CCF}$ derived from the fitting curves are convincing? To
address this question, we compare the lags calculated from the
fitting curves with those derived from the observed data. Using the
observed data and the same CCF method, we calculate the lags
($\tau_{CCF}^{\ast}$) of GRBs 980425 and 060218 only in the X-ray
energy bands or in the gamma-ray energy bands. The errors of
$\tau_{CCF}^{\ast}$ are evaluated by the simulations. The results
are also reported in Table 3. From Table 3, we find the calculated
lags from the two methods are consistent, but the values of
$\tau_{CCF}^{\ast}$ estimated from the observed X-ray light curves
for GRB 980425 have large errors. Thus, our estimated lags from the
fitting curves are convincing.

Based on the above results, we can analyze the relationships between
the three types of lags. For the purpose of unified comparison, we
use all the quantities derived from the fitting light curves. The
plots of $\tau_{cen}$ vs. $\tau_{peak}$, $\tau_{cen}$ vs.
$\tau_{CCF}$, and $\tau_{CCF}$ vs. $\tau_{peak}$ are displayed in
Figure 5. The results of correlation analysis for the three
quantities are listed in Table 2. We find that $\tau_{CCF}$ and
$\tau_{peak}$ are highly correlated for the multi-wavelength
observations in GRBs 980425 and 060218, while the other pairs of the
quantities are less well correlated. In addition, we find that
$\tau_{cen}$ is systematically larger than both $\tau_{peak}$ and
$\tau_{CCF}$. These results are well consistent with those derived
in typical LLWP-GRBs (Z07). As suggested by Z07, $\tau_{CCF}$ is
mainly caused by a shifting of the pulse peaks, while $\tau_{cen}$
is not. We suspect that $\tau_{CCF}$ and $\tau_{cen}$ reflect
different aspects of pulse evolution, with one representing the
shifting of the pulse peaks and the other describing an enhancement
of the pulse time scales. Under this interpretation, the lag caused
by the stretching of pulses is always larger than that caused by the
shifting of the pulse peaks. In addition, the nonlinear fluctuations
statistically present between the different types of lag
measurements (e.g., Figure 5), which might be due to the process of
pulse evolution and/or to instrumental response. In other words,
each type of lag measurement may be well sensitive to different
variations pertaining to pulse evolution; these variations may
depend upon pulse shape, energy, and/or signal-to-noise. It is
possible that the different types of lag measurements could be used
as a tool for probing aspects of pulse evolution. Thus, we propose
to reveal the evolution of a pulse in detail, both the pulse peak
lag and the centroid lag should be measured.

Recently, Lu et al. (2006) considered the contributions of the
curvature effect of fireballs to the spectral lag (see also Shen et
al. 2005), and tentatively studied the dependence of spectral lag on
energy. They considered a wide energy band ranging from 0.2 to 8000
keV, and then divided the band into 14 geometrical uniform energy
bands: $0.2-0.4, 0.5-1, 1-2, 2-4, 5-10, 10-20, 20-40, 50-100,
100-200, 200-400, 500-1000, 1000-2000, 2000-4000$ and $4000-8000$
keV. Subsequently, they measured the spectral lags\footnote{Note
that the spectral lag of the Lu et al. (2006) paper was defined as
the time between the peaks of the light curves in two different
energy bands, which is the pulse peak lag in this paper.} between
the first energy band ($0.2-0.4$ keV) and any of the other bands and
pointed out that the lags increases with energy following the law of
$\tau\propto E$, and then saturates at a certain energy [see the
left panel of Fig. 13 in Lu et al. (2006)]. Motivated by this, we
also investigate the dependence of the three types of lags on energy
for GRBs 980425 and 060218. We analyze the lags between the lowest
energy band ($2-5$ keV for GRB 980425 and $0.3-2$ keV for GRB
060218) and all of the other higher energy bands as performed by Lu
et al. (2006). Figure 6 shows the relationship between $\tau$ and
$E$ (here $E$ denotes the energy of the corresponding high-energy
band). We find from Figure 6, the three types of lags relative to
the same low-energy band increase with the energy of the
corresponding high-energy band, but their increases become shallow
at higher energies. The trend of the $\tau- E$ relation for the two
bursts seems to be similar with that obtained by Lu et al. (2006).
Probably, the curvature effect of the fireballs plays a role in
producing the relation (see, e.g., Qin et al. 2004, 2005), which is
currently not clear.

\section{Comparison with Typical Long-Lag, Wide-Pulse GRBs}

N05 analyzed the temporal and spectral behavior of the wide pulses
in 24 long-lag BATSE bursts and suggested that these events may form
a separate subclass of GRBs\footnote{GRB 980425 is included in the
N05 sample.}. Although GRBs 980425 and 060218 are two very peculiar
low-luminosity events, both of them have a simple temporal
structure, and their light curves are composed of a long duration
single-pulse with long spectral lag. It is very interesting to see
whether they have the different temporal properties with typical
LLWP-GRBs to explain their abnormal luminosities. In order to
clarify this issue, we first compare the distribution of ($w$,
$\tau$) of the two bursts with that of the bursts in the N05 sample.
Besides GRB 060218, the values of $w$ of other bursts are directly
taken from Table 2 of N05. The definition of $w$ given by N05 is the
width between the two $1/e$ peak intensity points of pulse, we also
measure the pulse width of GRB 060218 in the different energy bands
according to this definition. We obtain $w=1053\pm275$, $1574\pm68$,
$2107\pm73$ and $3668\pm214$ s in the energy bands $15-150$, $5-10$,
$2-5$ and $0.3-2$ keV, respectively. In general, the spectral lags
($\tau_{B31}$) of the BATSE bursts between energy bands $100-300$
keV ($B3$) and $25-50$ keV ($B1$) could be well estimated and widely
adopted by many authors. We only analyze the spectral lags in the
two energy bands\footnote{The data are taken from Z07 for the N05
sample.}. The values of $\tau_{peak,B31}$ ($177\pm16$ s) and
$\tau_{CCF,B31}$ ($61\pm26$ s) of GRB 060218 estimated by L06 are
available. Using the same extrapolated method as done by L06, we
obtain $\tau_{cen,B31}=219\pm30$ s, $w_{B1}=1065\pm61$ s and
$w_{B3}=585\pm34$ s for GRB 060218. Figure 7 shows the relationships
of $\tau_{peak,B31}$, $\tau_{CCF,B31}$ and $\tau_{cen,B31}$ against
$w_{B1}$ and $w_{B3}$ for the N05 sample as well as GRB 060218. We
find from Figure 7 that the distribution of ($w$, $\tau$) of GRB
980425 is completely consistent with that of the other LLWP-GRBs (as
pulse width increase, the spectral lag tends to increase, see N05),
and GRB 060218 also fall into the same sequence, although it has the
longest pulse width and spectral lag observed to date.

Recently, Peng et al. (2007) suggested that the correlation between
pulse spectral lag and pulse width might be caused by the Lorentz
factor of the GRBs. However, the pulse relative spectral lag (RSL),
which is defined as the ratio of the pulse spectral lag between
light curves observed in two different energy bands (in general, the
BATSE $B1$ and $B3$ bands are adopted) to the pulse width (see,
Zhang et al. 2006a, 2006b, Peng et al. 2007; Zhang \& Xie 2007), is
an unique and intrinsic quantity since such definition can reduce
both Doppler and cosmological time dilation effects on the
observations owing to $\tau\propto\Gamma^{-2}\propto(1 + z)$ and
$w\propto\Gamma^{-2}\propto(1 + z)$ (Zhang et al. 2006b; Norris et
al. 2000; Kocevski \& Liang 2006; Peng et al. 2007; Zhang \& Xie
2007). Therefore, we also analyze the relation between the pulse
RSLs and pulse widths for the typical LLWP-GRBs as well as GRBs
980425 and 060218. The results are plotted in Figure 8. We find from
Figure 8 that the pulse RSLs are not correlated with the pulse
widths, and the pulse RSLs of GRBs 980425 and 060218 are fully
consistent with those of the other LLWP-GRBs.

In addition, we also compare the two bursts with the events of the
N05 sample in the panel of $r/d$ vs. $w$. Using the data of Table 2
and the equation (5) in N05, we derive the values of $r/d$ in the
$B1$ and $B3$ energy bands for all the bursts in the N05 sample.
Figure 9 shows the plots of $r/d$ vs. $w$ in the $B1$ and $B3$
energy bands\footnote{The values of $r/d$ for GRB 060218 have large
errors at higher energies, and which can not be estimated well, so
we take the value in the $15-150$ keV ($0.56\pm0.15$) as that in the
$B1$ and $B3$ energy bands, which can not affect the results more.}.
As can be seen from Figure 9 that the pulse rise-to-decay ratios of
GRBs 980425 and 060218 are in good agreement with those of the other
LLWP-GRBs. These results indicate that GRBs 980425 and 060218 may
share the similar radiation physics with them.

\section{Conclusions and Discussion}

We have analyzed the prompt lightcurve characteristics of GRBs
980425 and 060218 from X-ray to gamma-ray energy bands. We find that
both the pulse width $w$ and the ratio of pulse rise-to-decay $r/d$
are energy-dependent for these two bursts over a broad energy band.
There exists a significant trend that the pulses of these two bursts
tend to be narrower and more symmetry with respect to the higher
energy bands. Both the X-rays and gamma-rays of the two events
follow the same $w - E$ and $r/d - E$ relations, but the power-law
indices of the $w - E$ relations are somewhat larger than those
observed previously in typical GRBs (Fenimore et al. 1995; Norris et
al. 1996; N05; Peng et al. 2006).

The light curves of GRBs 980425 and 060218 show significant spectral
lags, with the pulse peaks shifting later time at lower energies. We
calculate the three types of lags $\tau_{peak}$, $\tau_{cen}$ and
$\tau_{CCF}$, with the pulse peaking time ($t_{peak}$), the pulse
centroid time ($t_{cen}$), and the cross-correlation function (CCF).
The derived $t_{peak}$ and $t_{cen}$ are a power-law function of
energy, and $\tau_{CCF}$ is strongly correlated with $\tau_{peak}$,
but the other pairs of the quantities are less well correlated. Our
analysis also show that $\tau_{cen}$ is systematically larger than
both $\tau_{peak}$ and $\tau_{CCF}$. In addition, the relationships
between the three types of lags and energy are investigated as well.
We find that the lags relative to the same low-energy band increase
with the energy of the corresponding high-energy band, but their
increases becomes shallow at higher energies.

Although GRBs 980425 and 060218 are two very peculiar low-luminosity
events, the temporal and spectral characteristics of these two
bursts are normal when compared to other typical LLWP-GRBs.

Our analysis is performed in the observer frame, rather than in the
GRB rest frame. This makes the comparison slightly inappropriate,
since GRBs 980425 and 060218 are low-redshift bursts, but the normal
long-lag GRBs have been observed at larger redshifts (typically
$z\sim1$). Recently, Hakkila et al. (2008) found that the rest frame
pulse duration ($w_{0}$), pulse peak lag ($\tau_{0}$) and isotropic
pulse peak luminosity ($L$) are highly correlated for the pulses of
BATSE GRBs with known redshifts. Remarkably, the underluminous GRB
980425 follows the $w_{0}-\tau_{0}$ relation well, but deviates from
the $L-\tau_{0}$ relation. Meanwhile, we also analyze the
distribution of GRB 060218 in both the $w_{0}-\tau_{0}$ and
$L-\tau_{0}$ panels (see Fig. 10). From Figure 10 we find that GRB
060218 also complies with the $w_{0}-\tau_{0}$ relation well, and it
is inconsistent with the $L-\tau_{0}$ correlation. This result
further reinforces our conclusion that the temporal and spectral
characteristics of GRBs 980425 and 060218 are normal. In addition,
besides the time dilation effect on the rest frame lags and
durations (the correction is $(1+z)^{-1}$) has been widely
concerned, the energy correction (K correction) also affected the
two rest frame quantities, since the normal pulses are subsequently
observed at lower energies than those of the low-luminosity pulses
and both the lags and durations are energy-dependent ($\tau\propto
E^{-0.4}$ and $w\propto E^{-0.4}$, see, e.g., Norris et al. 1996;
N05; Z07). The energy correction to the rest frame for the lags and
durations is approximately $(1+z)^{0.33}$ (e.g., Gehrels et al.
2006). This effect is not considered here. When comparing these
observations to observer frame observations of higher-z bursts, both
the energy correction and time dilation correction should be taken
into account.

Stern et al. (1999) first suggested that there is a group of
¡°simple¡± bursts with peak fluxes near the BATSE trigger threshold:
the average profile of dim bursts were less complex than that of
bright bursts. Norris (2002) found that the proportion of long-lag
bursts within long-duration bursts increases from negligible among
bright BATSE bursts to $\sim 50 \%$ at the trigger threshold. N05
proposed that these long-lag bursts may be underluminous and form a
separate subclass of GRBs (see also, Liang et al. 2007; Le \& Dermer
2007; Guetta \& Della Valle 2007; Ghisellini et al. 2007; Daigne \&
Mochkovitch 2007). However, the Hakkila et al. (2008) results
challenge this statement. They found that $L$, $\tau_{0}$ and
$w_{0}$ are correlated intrinsic properties of most GRB pulses. GRBs
980425 and 060218 are fully consistent with the $w_{0}-\tau_{0}$
relation holding for the normal GRB pulses. Given this, the evidence
for a separate class of LLWP-GRBs seems to be much weaker. However,
both of them are two apparent outliers with respect to the
$L-\tau_{0}$ relation. This result makes the underluminous features
of GRBs 980425 and 060218 that much more unusual. Based on the fact
that redshifts of three such bursts are available [GRB 980425,
Galama et al. 1998; 031203, Malesani et al. 2004 and 060218, Mirabal
et al. 2006], some authors argued that the low-luminosity bursts are
probably relatively nearby, and the local event rate of these events
should be much higher than that expected from the high-luminosity
GRBs (Cobb et al. 2006; Pian et al. 2006; Soderberg et al. 2006;
Liang et al. 2007; Le \& Dermer 2007; Guetta \& Della Valle 2007).
There are two scenarios which were proposed to explain their
wide-pulse, long-lag, and underluminous features. One possible
scenario is that these GRBs are normal events viewed off-axis (e.g.,
Nakamura 1999; Salmonson 2000; Yamazaki et al. 2003). The second
scenario is that these features are intrinsic, may be due to their
lower Lorentz factors (Kulkarni et al. 1998; Woosley \& MacFadyen
1999; Salmonson 2000; Dai et al. 2006; Wang et al. 2007) or a
different type of central engine (e.g., neutron stars rather than
black holes; see references, Mazzali et al. 2006; Soderberg et al.
2006; Toma et al. 2007).

\textbf{Acknowledgments}

I would like to address my great thanks to the anonymous referee for
his/her helpful comments and suggestions which helped me to improve
the paper greatly. I also thank En-Wei Liang, Jin-Ming Bai, Yi-Ping
Qin and Bin-Bin Zhang for their helpful discussions. I am grateful
to Prof. Jon Hakkila for providing the pulse duration and pulse lag
data of BATSE bursts with known redshifts. I also express our thanks
to Dr. Alok Gupta (ARIES, India) for going through the paper and
making several suggestions to improve the language. This work is
supported by National Natural Science Foundation of China (No.
10573030 and No. 10533050).

% \appendix

% \section{}
% \label{}

% Bibliographic references with the natbib package:
% Parenthetical: \citep{Bai92} produces (Bailyn 1992).
% Textual: \citet{Bai95} produces Bailyn et al. (1995).
% An affix and part of a reference:
%   \citep[e.g.][Ch. 2]{Bar76}
%   produces (e.g. Barnes et al. 1976, Ch. 2).

%\clearpage

%% the file, uncomment the \plotone, \plottwo, and
%% If you need a layout that cannot be achieved with \plotone or
%% \plottwo, you can invoke the graphicx package directly with the
%% \includegraphics command or use \plotfiddle. For more information,
%% please see the tutorial on "Using Electronic Art with AASTeX" in the
%% documentation section at the AASTeX Web site,
%% http://www.journals.uchicago.edu/AAS/AASTeX.
%%
%% The examples below also include sample markup for submission of
%% supplemental electronic materials. As always, be sure to check
%% the instructions to authors for the journal you are submitting to
%% for specific submissions guidelines as they vary from

%% This example uses \plotone to include an EPS file scaled to
%% 80% of its natural size with \epsscale. Its caption
%% has been written to indicate that additional figure parts will be
%% available in the electronic journal.

\begin{figure}
\epsscale{.90} \plotone{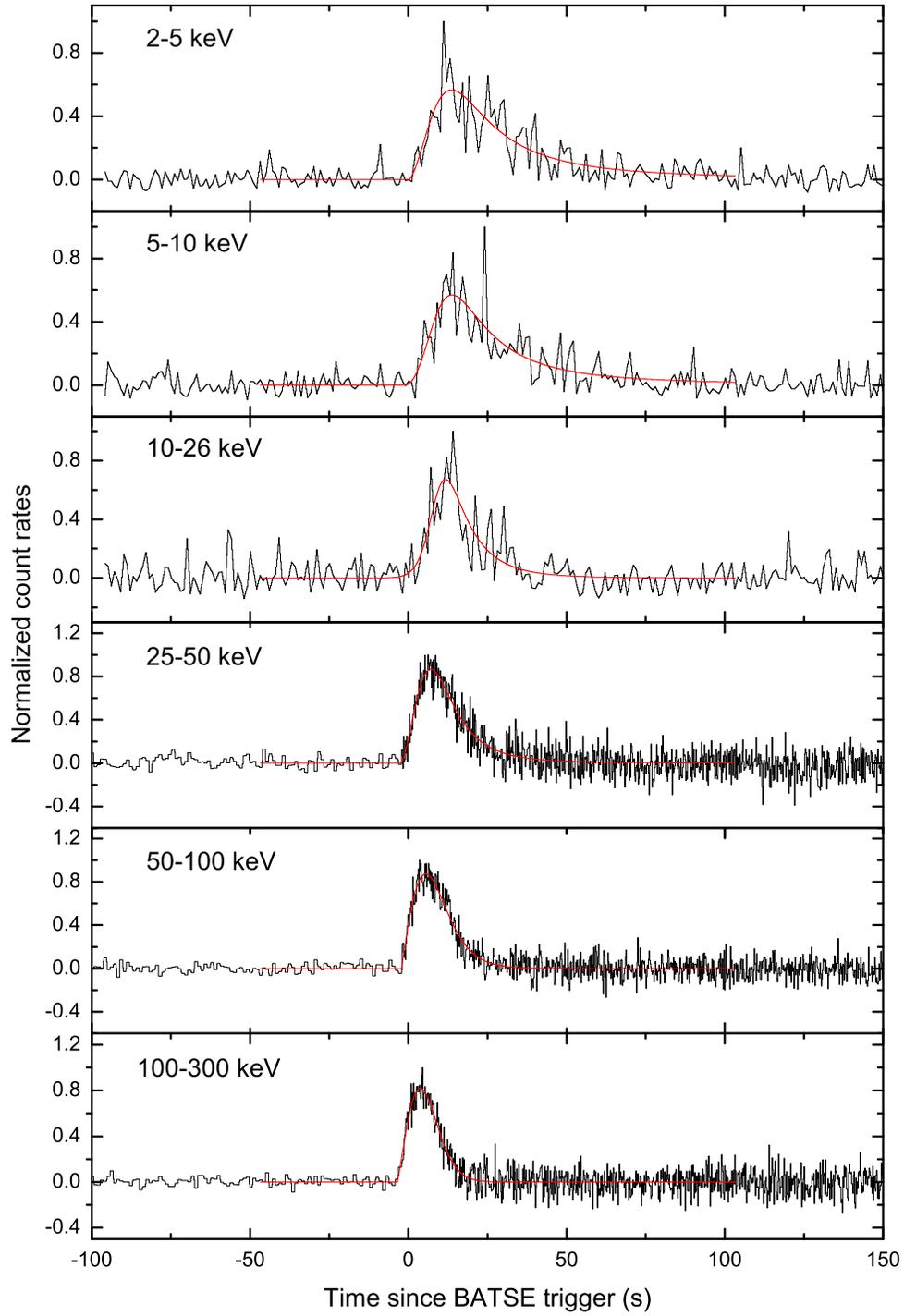} \caption{BeppoSAX and BATSE light
curves of GRB 980425. The count rates have been normalized to the
peak of each light curve. The fitting curves with eq. (1) are
plotted. \label{fig1}}
\end{figure}

%\clearpage

\begin{figure}
\epsscale{.90} \plotone{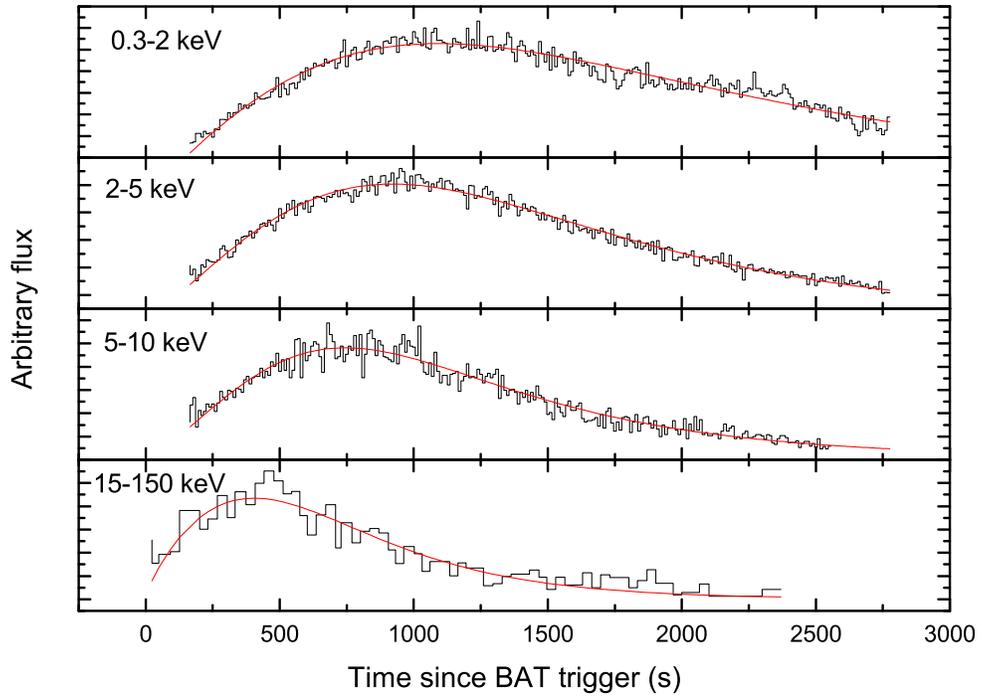} \caption{XRT and BAT light curves of
the nonthermal emission of GRB 060218. The fitting curves with eq.
(1) are also plotted. The data are taken from L06. \label{fig2}}
\end{figure}

\begin{figure}
\epsscale{.90} \plotone{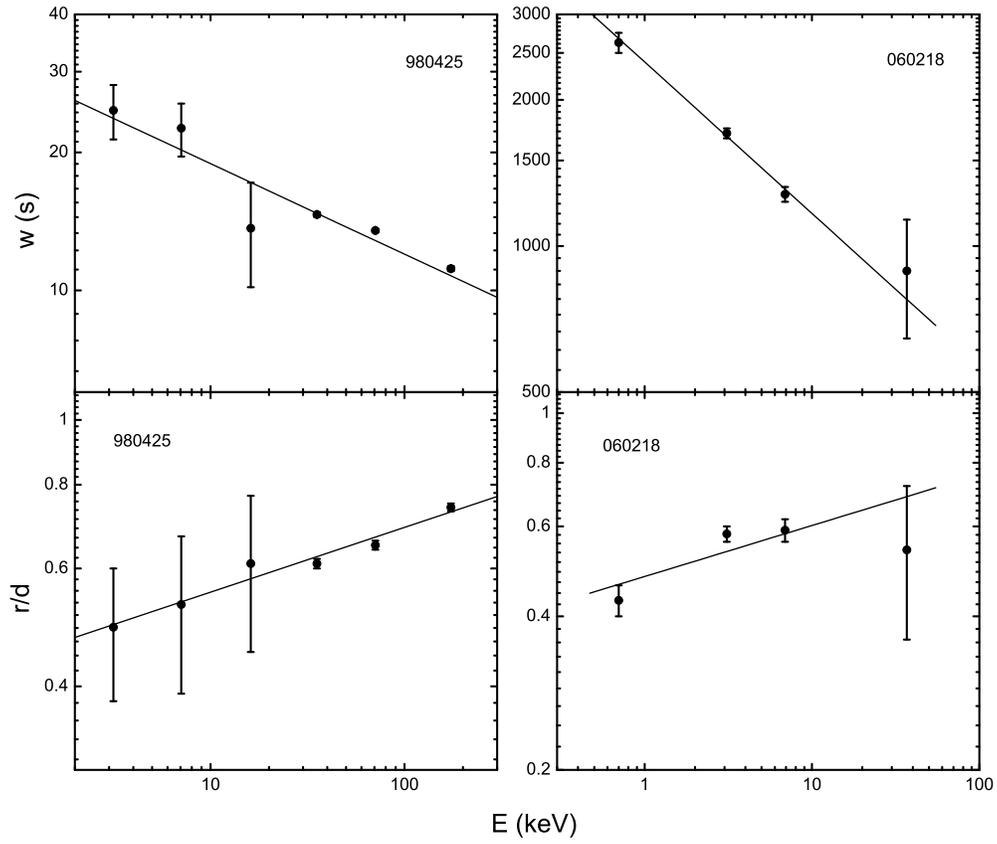} \caption{Dependence of the pulse
width ($top$) and pulse rise-to-decay ratio ($bottom$) on energy in
GRBs 980425 and 060218. The solid lines in the plots are the best
fits.\label{fig3}}
\end{figure}

\begin{figure}
\epsscale{.90} \plotone{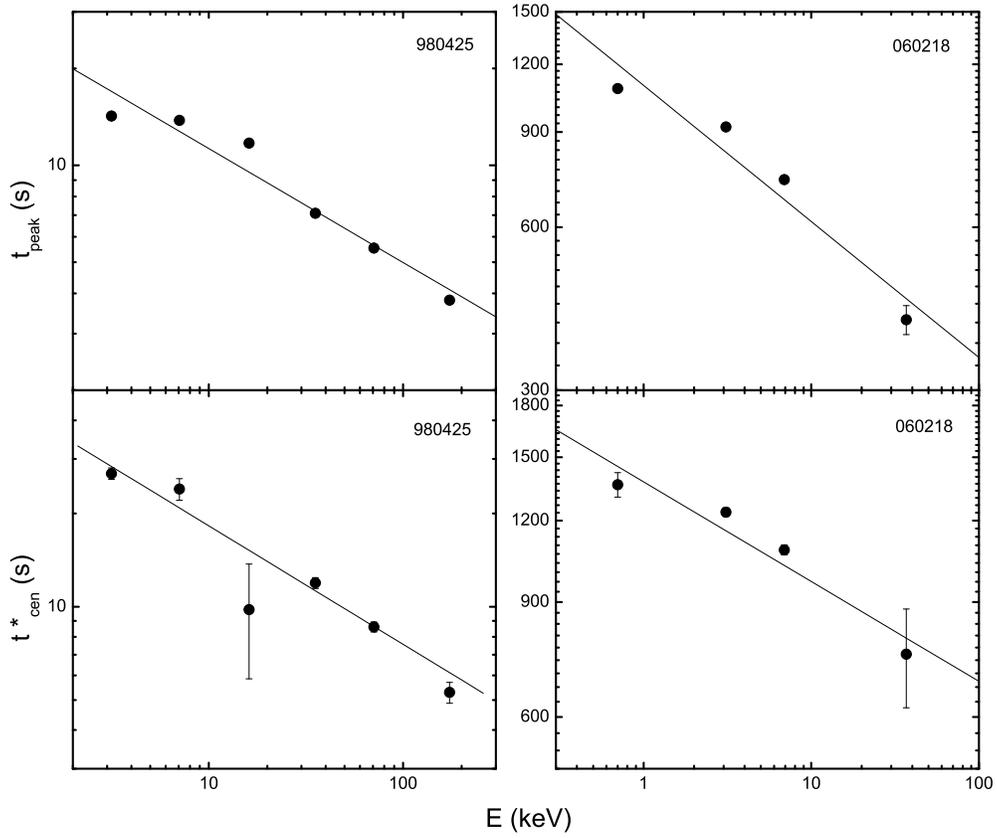} \caption{The plots of the pulse peak
time against energy ($top$) and the centroid time versus energy
($bottom$) in GRBs 980425 and 060218. The $t_{cen}^{\ast}$ are
estimated directly based on the observed data. The solid lines
represent the best fits.\label{fig4}}
\end{figure}

\begin{figure} \epsscale{.90} \plotone{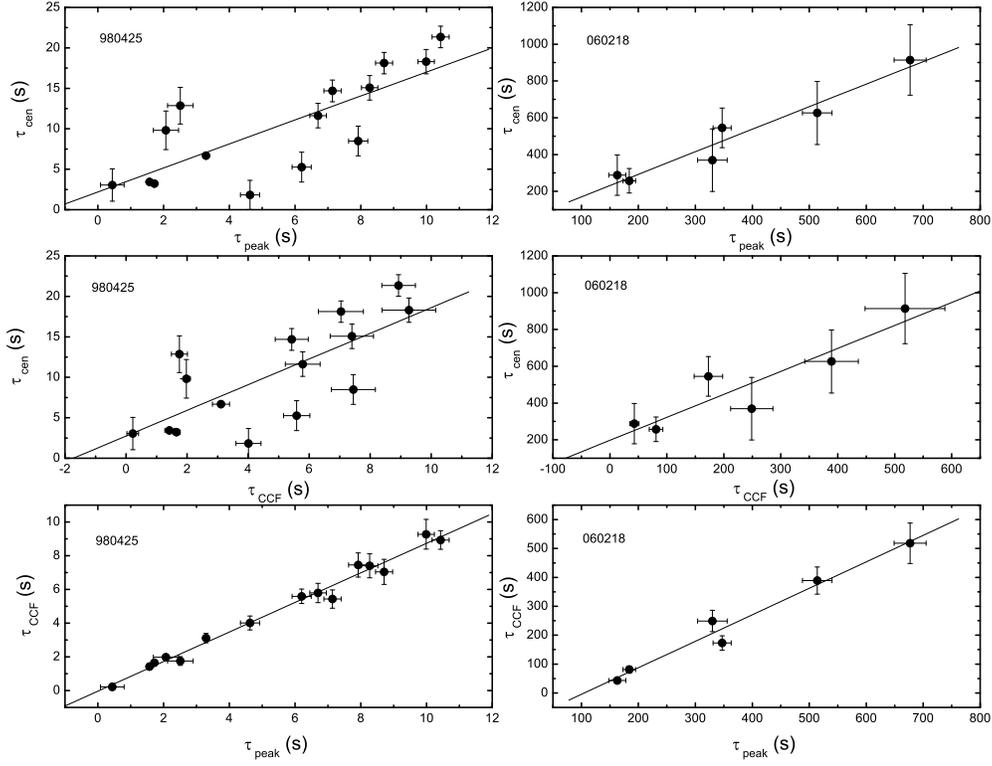} \caption{Relationships
between the three types of lags ($\tau_{cen}$, $\tau_{peak}$, and
$\tau_{CCF}$). The solid lines are the regression lines, where the
correlation coefficients from the top to bottom panels are 0.78,
0.74, and 0.99 for GRB 980425, and 0.97, 0.92 and 0.99 for GRB
060218, respectively.\label{fig5}}
\end{figure}

\begin{figure}
\epsscale{.90} \plotone{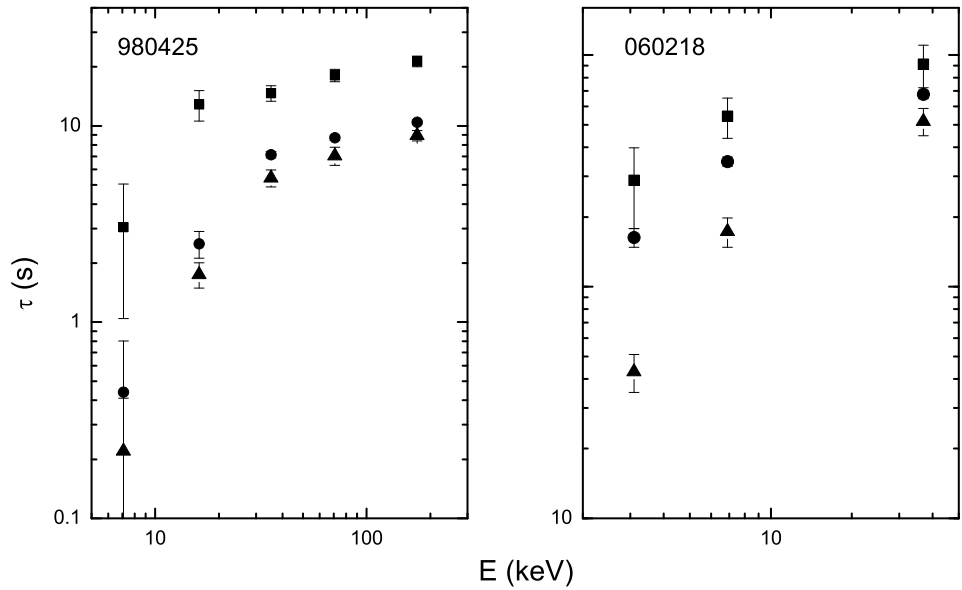} \caption{plots of $\tau$ vs. $E$,
where $\tau$ are spectral lags between the first energy band ($2-5$
keV for GRB 980425 and $0.3-2$ keV for GRB 060218) and any of the
other high-energy bands, $E$ is the energy of the corresponding
high-energy band. The circle, square and triangle symbols represent
the lags derived from the pulse peak time, centroid time and CCF,
respectively. \label{fig6}}
\end{figure}

\begin{figure}
\epsscale{.90} \plotone{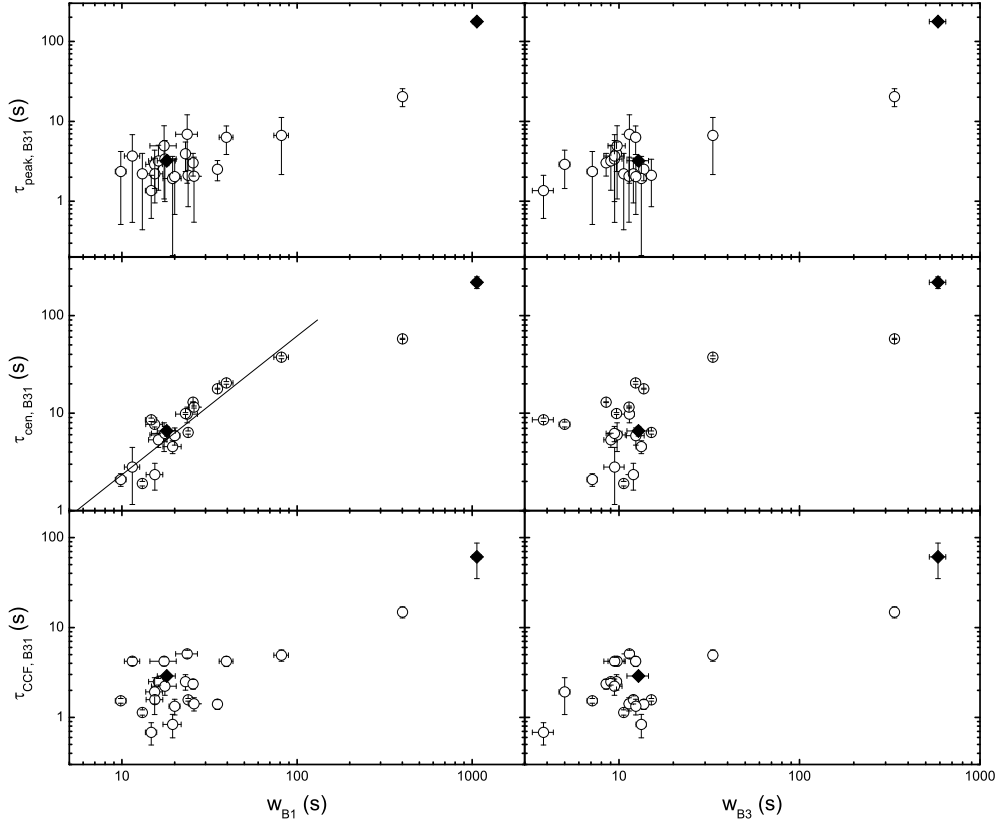} \caption{Relation between pulse
spectral lags and pulse widths, where $\tau_{peak,B31}$,
$\tau_{cen,B31}$ and $\tau_{CCF,B31}$ are the pulse peak lags,
centroid lags and CCF lags in the $100-300$ keV ($B3$) and $25-50$
keV ($B1$) bands, $w_{B1}$ and $w_{B3}$ are the pulse width measured
between the two $1/e$ intensity points defined by N05 in the $B1$
and $B3$ bands, respectively. The solid line is the best fit
($\tau_{cen,B31}\approx0.089w_{B1}^{1.42}$ s) obtained by N05. The
filled diamonds represent GRB 980425 and GRB 060218, and the open
circles are the other bursts in the N05 sample besides GRB
980425.\label{fig7}}
\end{figure}

\begin{figure}
\epsscale{.95} \plotone{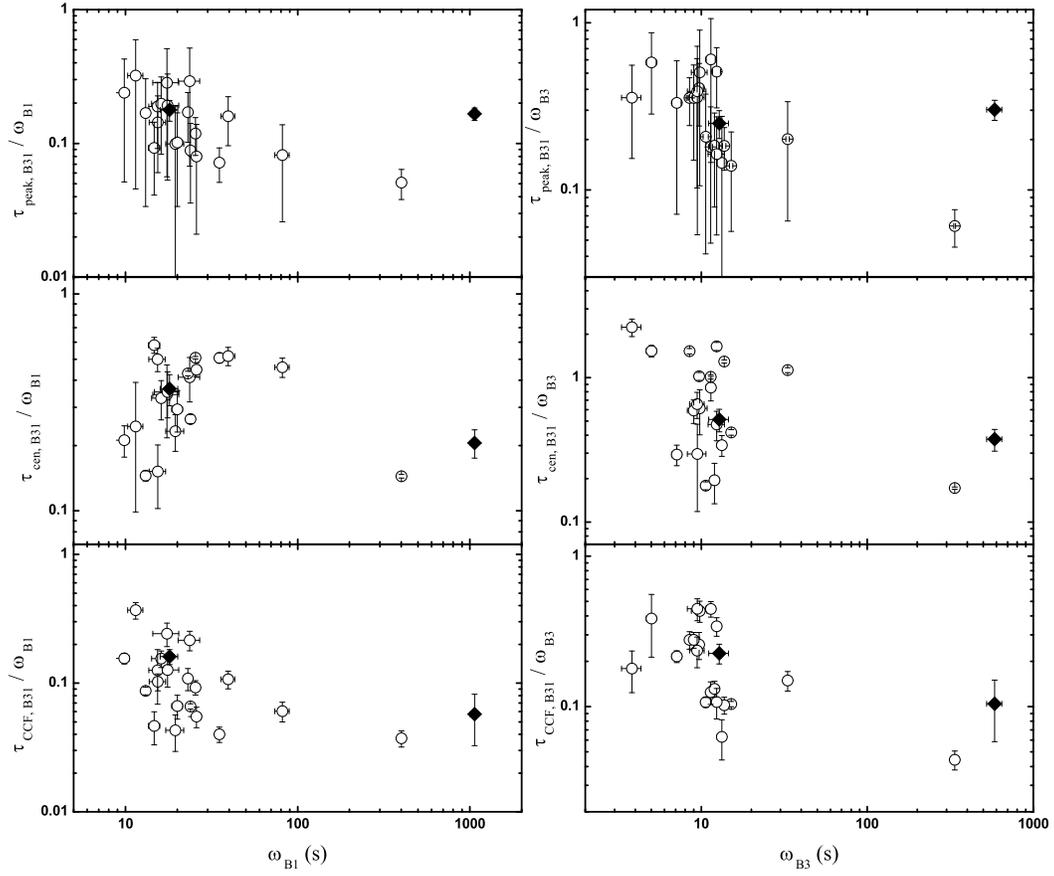} \caption{Pulse relative spectral
lags vs. pulse widths. The other symbols are same as Fig. 7.
\label{fig8}}
\end{figure}

\begin{figure}
\epsscale{.80} \plotone{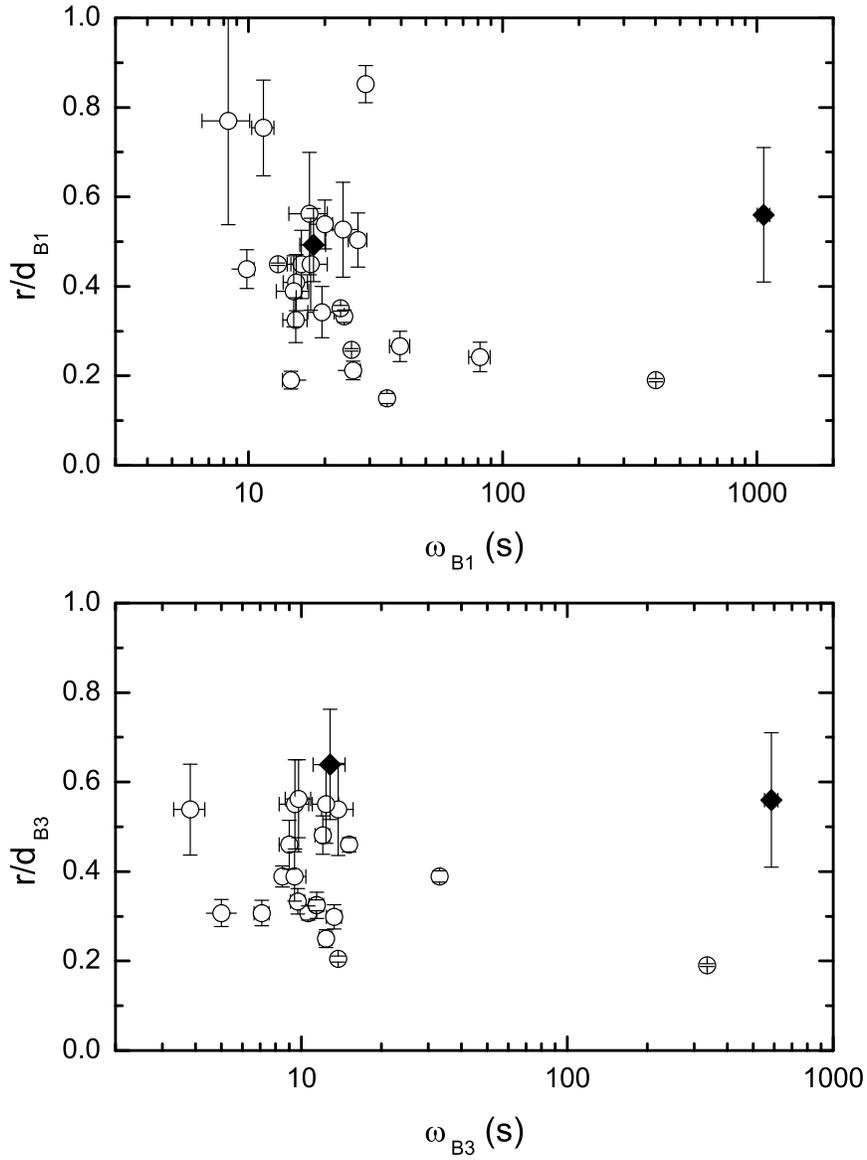} \caption{Pulse rise-to-decay ratios
vs. widths. The $r/d_{B1}$ and $r/d_{B3}$ are the pulse
rise-to-decay ratio measured in the $B1$ and $B3$ bands. The other
symbols are same as Fig. 7. \label{fig9}}
\end{figure}

\begin{figure}
\epsscale{.90} \plotone{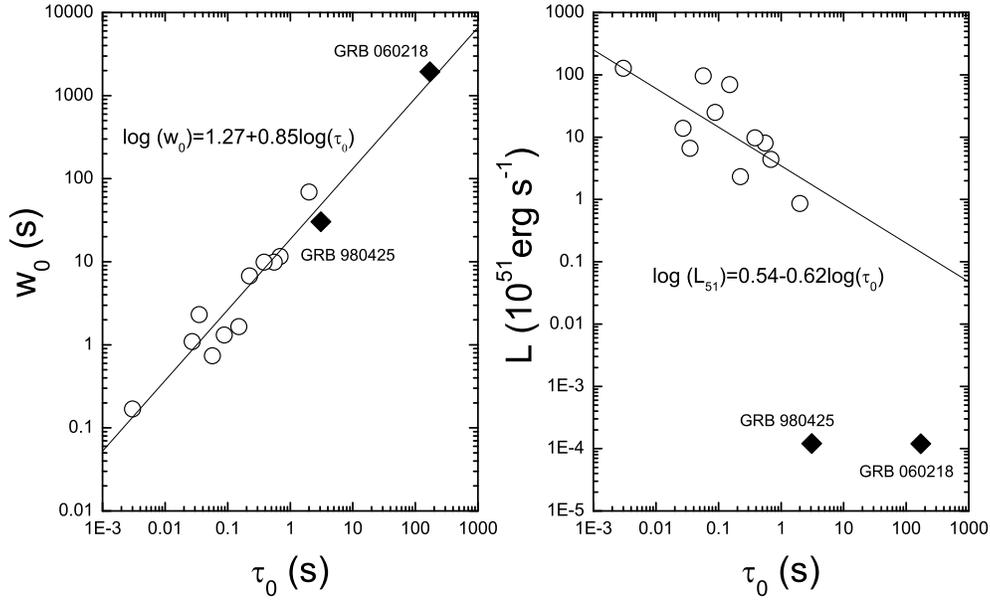} \caption{\emph{Left}: Rest frame
pulse duration $w_{0}$ vs. pulse peak lag $\tau_{0}$ for fit pulses
of BATSE GRBs having known redshifts (the data are taken from
Hakkila et al. 2008) as well as GRB 060218. \emph{Right}: Isotropic
pulse peak luminosity $L$ vs. pulse peak lag $\tau_{0}$ for the
pulses shown in the left panel. The open circles represent the
pulses from GRB 971214, GRB 980703, GRB 970508, GRB 990510, GRB
991216 and GRB 990123, and the filled diamonds represent GRB 980425
and GRB 060218. The solid lines are the best fits obtained by
Hakkila et al. (2008). \label{fig10}}
\end{figure}

%% Here we use \plottwo to present two versions of the same figure,
%% one in black and white for print the other in RGB color
%% for online presentation. Note that the caption indicates
%% that a color version of the figure will be available online.
%%

\clearpage

\begin{deluxetable}{lllllll}
\tablewidth{0pt} \tablecaption{Broadband temporal characteristics of
GRB 980425 and GRB 060218}

\tablehead{
 \colhead{ Band} & \colhead{$t_{peak}$ }&\colhead{$t_{cen}^{\ast}$}& \colhead{$t_{cen}$}& \colhead{$w$}
&\colhead{$r/d$}  &\colhead{$E$}\\
 \colhead{(keV)} & \colhead{(s)}&\colhead{(s)}&\colhead{(s)}& \colhead{(s)}
&\colhead{} &\colhead{(keV)}}

\startdata
             &               &               &GRB 980425               & & &\\
\hline
 (1) 2-5     &14.24$\pm$0.26 &26.91$\pm$1.14 &27.10$\pm$1.32 &24.71$\pm$3.36 &0.49$\pm$0.11 &3.2\\
 (2) 5-10    &13.80$\pm$0.25 &23.98$\pm$1.93 &24.05$\pm$1.51 &22.58$\pm$2.98 &0.53$\pm$0.14 &7.1\\
 (3) 10-26   &11.73$\pm$0.29 & 9.79$\pm$3.94 &14.24$\pm$1.84 &13.67$\pm$3.51 &0.61$\pm$0.16 &16.1\\
 (4) 25-50   & 7.10$\pm$0.01 &11.94$\pm$0.48 &12.42$\pm$0.18 &14.64$\pm$0.17 &0.61$\pm$0.01 &35.4\\
 (5) 50-100  & 5.53$\pm$0.01 & 8.61$\pm$0.32 & 8.98$\pm$0.07 &13.51$\pm$0.11 &0.65$\pm$0.01 &70.7\\
 (6) 100-300 & 3.81$\pm$0.01 & 5.29$\pm$0.41 & 5.75$\pm$0.08 &11.16$\pm$0.14 &0.74$\pm$0.01 &173.2\\
\hline
             &               &               &GRB 060218$^{\star}$             & & &\\
\hline
 (1) 0.3-2  &1082$\pm$13 &1362$\pm$59 &1687$\pm$98 &2625$\pm$125 &0.43$\pm$0.03 &0.7\\
 (2) 2-5    &  919$\pm$7 &1236$\pm$19 &1399$\pm$49 & 1707$\pm$40 &0.58$\pm$0.02 &3.1\\
 (3) 5-10   &  735$\pm$9 &1082$\pm$19 &1142$\pm$45 & 1278$\pm$45 &0.59$\pm$0.03 &6.9\\
 (4) 15-150 & 405$\pm$25 &749$\pm$129 &773$\pm$164 & 889$\pm$244 &0.54$\pm$0.18 &36.9\\

\enddata
\tablenotetext{\ast}{ The values of $t_{cen}^{\ast}$ are estimated
directly based on the observed data.}

\tablenotetext{\star}{ The values of $t_{peak}$, $w$, $r/d$ and $E$
of GRB 060218 are taken from L06.}
\end{deluxetable}

\begin{deluxetable}{ll}
\tablewidth{0pt} \tablecaption{Correlations of the temporal
structures of GRB 980425 and GRB 060218}

\tablehead{\colhead{GRB 980425}& \colhead{GRB 060218}  }

\startdata
log $w=(1.47\pm0.06)-(0.20\pm0.04)$ log $E$        & log $w=(3.38\pm0.02)-(0.31\pm0.03)$ log $E$    \\
log $(r/d)=(-0.35\pm0.01)+(0.10\pm0.01)$ log $E$   & log $(r/d)=(-0.32\pm0.03)+(0.10\pm0.03)$ log $E$ \\
log $t_{peak}=(1.41\pm0.07)-(0.35\pm0.04)$ log $E$ & log $t_{peak}=(3.04\pm0.04)-(0.25\pm0.05)$ log $E$\\
log $t_{cen}^{\ast}=(1.63\pm0.10)-(0.40\pm0.07)$ log $E$ & log $t_{cen}^{\ast}=(3.14\pm0.03)-(0.15\pm0.03)$ log $E$\\
log $t_{cen}=(1.66\pm0.04)-(0.39\pm0.03)$ log $E$ &log $t_{cen}=(3.22\pm0.02)-(0.20\pm0.02)$ log $E$\\
$\tau_{cen}=(2.18\pm2.09)+(1.48\pm0.33)\tau_{peak}$ &
$\tau_{cen}=(47\pm62)+(1.23\pm0.15)\tau_{peak}$ \\
$\tau_{cen}=(2.74\pm2.23)+(1.59\pm0.40)\tau_{CCF}$ &
$\tau_{cen}=(197\pm79)+(1.25\pm0.27)\tau_{CCF}$ \\
$\tau_{CCF}=(0.11\pm0.12)+(0.86\pm0.04)\tau_{peak}$ &
$\tau_{CCF}=(-100\pm17)+(0.91\pm0.08)\tau_{peak}$ \\
\enddata
\end{deluxetable}

\begin{deluxetable}{llllll}
\tablewidth{0pt} \tablecaption{Multi-band spectral lags of GRBs
980425 and 060218}

\tablehead{
 \colhead{Bands} & \colhead{$\tau_{peak}$}& \colhead{$\tau_{cen}^{\ast}$} & \colhead{$\tau_{cen}$}
 & \colhead{$\tau_{CCF}$} & \colhead{$\tau_{CCF}^{\ast}$}  \\
  & \colhead{(s)}&\colhead{(s)}& \colhead{(s)} & \colhead{(s)}& \colhead{(s)}}

\startdata
             &               &               &GRB 980425               & & \\
\hline
 (1)-(2)  & 0.44$\pm$0.36 & 2.93$\pm$2.24 & 3.05$\pm$2.01 &0.22$\pm$0.19 &0.45$\pm$1.56\\
 (1)-(3)  & 2.51$\pm$0.39 &17.12$\pm$4.10 &12.86$\pm$2.26 &1.75$\pm$0.26 &1.87$\pm$1.69\\
 (1)-(4)  & 7.14$\pm$0.26 &14.97$\pm$1.23 &14.68$\pm$1.33 &5.43$\pm$0.54 &...          \\
 (1)-(5)  & 8.71$\pm$0.26 &18.30$\pm$1.18 &18.12$\pm$1.32 &7.04$\pm$0.74 &...          \\
 (1)-(6)  &10.43$\pm$0.26 &21.62$\pm$1.21 &21.35$\pm$1.32 &8.93$\pm$0.55 &...          \\
 (2)-(3)  & 2.07$\pm$0.38 &14.19$\pm$4.39 & 9.81$\pm$2.38 &1.98$\pm$0.13 &1.45$\pm$1.15\\
 (2)-(4)  & 6.70$\pm$0.25 &12.04$\pm$1.99 &11.63$\pm$1.52 &5.79$\pm$0.57 &...          \\
 (2)-(5)  & 8.27$\pm$0.25 &15.37$\pm$1.96 &15.07$\pm$1.51 &7.40$\pm$0.71 &...          \\
 (2)-(6)  & 9.99$\pm$0.25 &18.69$\pm$1.97 &18.30$\pm$1.51 &9.27$\pm$0.88 &...          \\
 (3)-(4)  & 4.63$\pm$0.29 &-2.15$\pm$3.97 & 1.82$\pm$1.85 &4.01$\pm$0.41 &...          \\
 (3)-(5)  & 6.20$\pm$0.29 & 1.18$\pm$3.95 & 5.26$\pm$1.84 &5.59$\pm$0.43 &...          \\
 (3)-(6)  & 7.92$\pm$0.29 & 4.50$\pm$3.96 & 8.49$\pm$1.84 &7.45$\pm$0.72 &...          \\
 (4)-(5)  & 1.57$\pm$0.01 & 3.33$\pm$0.58 & 3.44$\pm$1.19 &1.42$\pm$0.15 &1.46$\pm$0.18\\
 (4)-(6)  & 3.29$\pm$0.01 & 6.65$\pm$0.63 & 6.67$\pm$0.20 &3.11$\pm$0.28 &3.08$\pm$0.32\\
 (5)-(6)  & 1.72$\pm$0.01 & 3.32$\pm$0.52 & 3.23$\pm$0.11 &1.65$\pm$0.12 &1.72$\pm$0.18\\
\hline
             &               &               &GRB 060218$^{\star}$               & & \\
\hline
 (1)-(2)  & 163$\pm$15 & 126$\pm$62 & 288$\pm$109 & 43$\pm$8  & 39$\pm$15\\
 (1)-(3)  & 347$\pm$16 &280$\pm$62  & 545$\pm$108 &173$\pm$25 &183$\pm$37\\
 (1)-(4)  & 677$\pm$28 &613$\pm$141 & 914$\pm$191 &518$\pm$70 &...       \\
 (2)-(3)  & 184$\pm$11 &154$\pm$27  & 257$\pm$66  & 81$\pm$12 & 71$\pm$12\\
 (2)-(4)  & 514$\pm$26 &487$\pm$130 & 626$\pm$171 &389$\pm$47 &...       \\
 (3)-(4)  & 330$\pm$26 &333$\pm$130 & 369$\pm$170 &249$\pm$37 &...       \\

\enddata
\tablenotetext{\ast}{The values of $\tau_{cen}^{\ast}$ and
$\tau_{CCF}^{\ast}$ are calculated directly based on the observed
data. }

\tablenotetext{\star}{The values of $\tau_{peak}$ and $\tau_{CCF}$
of GRB 060218 are taken from L06.}
\end{deluxetable}

\end{document}